\documentstyle{article}

\def\b{\begin{equation}}
\def\e{\end{equation}}

\def\ba{\begin{eqnarray}}
\def\ea{\end{eqnarray}}

\title{\bf Truncation-type methods and B\"acklund
transformations for ordinary
differential
equations: the third and fifth Painlev\'e equations}

\author{{\bf P. R. Gordoa,$^{\dagger}$ N. Joshi$^{\ddagger}$ and A.
Pickering$^{\dagger}$} \\ \\
             $^{\dagger}$ Area de Fisica Te\'orica \\
             Facultad de Ciencias \\
             Edificio de Fisica \\
             Universidad de Salamanca \\
             37008 Salamanca, Spain \\ \\
             $^{\ddagger}$ Department of Pure Mathematics \\
	University of Adelaide \\
             Adelaide \\ Australia 5005}

\date{}

\begin{document}

\maketitle

\thispagestyle{empty}

\begin{tabbing}

\noindent \smallskip
{\em Short title:}
\=  B\"acklund transformations for $P_{III}$ and $P_V$\\


Ordinary differential equations
% 05.45.Yv (solitons),

\end{tabbing}

\begin{abstract}

In a recent paper we presented a truncation-type method of deriving B\"acklund
transformations for ordinary differential equations. This method is based on a
consideration of truncation as a mapping that preserves the locations of a
natural subset of the movable poles that the equation possesses. Here we apply
this approach to the third and fifth Painlev\'e equations. For the third
Painlev\'e
equation we are able to obtain all fundamental B\"acklund transformations for
the
case where the parameters satisfy $\gamma\delta\neq0$. For the fifth Painlev\'e
equation our approach yields what appears to be all known B\"acklund
transformations.

\end{abstract}

\vfill

\hfill 14 January 2000

\newpage

\section{Introduction}

In a recent paper \cite{GJP99} we introduced a new approach to finding
B\"acklund
transformations for ordinary differential equations (ODEs). This
truncation-type
method constituted an extension to ODEs of an approach that had been
successfully
developed for partial differential equations (PDEs) \cite{Weiss}. The main idea
in \cite{GJP99} is to consider truncation as a mapping that preserves the
locations
of a natural subset of movable singularities. The generic solution of each of
the
Painlev\'e equations (except $P_I$) has pairs of movable simple poles with
leading
order coefficients of opposite sign. Thus the set of all poles of a solution
$y(x)$
decomposes into the union of two nonintersecting subsets $\cal P_+$ and $\cal
P_-$,
where $\cal P_+$ is the set of poles with positive choice of coefficient and
$\cal P_-$
is that with negative choice. In what follows we transform a generic solution
$y(x)$
of a Painlev\'e equation to a solution $Q(x)$ of the same equation, but with
possibly
different parameters, as
\b
y(x)=\rho(x)+Q(x),
\e
where we demand that $\rho(x)$ has poles at the elements of $\cal P_+$ and
$Q(x)$
has them at $\cal P_-$, or vice versa. We are then able to find B\"acklund
transformations
through a procedure that relies only on singularity analysis of the transformed
equation.

In \cite{GJP99} we applied this approach to $P_{II}$ and $P_{IV}$. In addition
to
obtaining B\"acklund transformations for these equations we also discussed
transformations
to related ODEs (the ODEs satisfied by $\rho(x)$), and also the application of
a
``double-singularity approach.'' In the present work we apply our approach to
$P_{III}$
and $P_{V}$, in order to obtain B\"acklund transformations. Consideration of
related
ODEs and the double-singularity approach for $P_{III}$ and $P_{V}$ will be
considered
elsewhere \cite{GJP00}. Descriptions of B\"acklund transformations for
$P_{III}$ and
$P_{V}$ can be found in \cite{FA,MCB,Gro}.

\section{B\"acklund transformations for $P_{III}$}

We take the third Painlev\'e equation in the form
\b
y''=\frac{y'^2}{y}-\frac{y'}{x}+\frac{\alpha
y^2+\beta}{x}+\gamma^2y^3-\frac{\delta^2}{y}
\label{p3y}
\e
where for reasons of convenience we have renamed two of the paramaters as
$\gamma^2$ and
$-\delta^2$ (conventionally labelled as $\gamma$ and $\delta$ respectively). A
generic
solution $y(x)$ of the third Painlev\'e equation is transformed to another
solution
$Q(x)$ of the same equation but with possibly different parameters $a, b, c$
and $d$ as
\b
y(x)=\rho(x)+Q(x).
\label{r+q}
\e
The dominant terms in the expression that results from the substitution of
equation
 (\ref{r+q}) into (\ref{p3y}) are, assuming $\gamma\neq 0$,
\b
\rho\rho''-\rho'^2 \approx \gamma^2 \rho^4,
\e
which can be integrated to give
\b
\rho'\approx \pm\gamma \rho^2.
\e
Taking first of all the minus sign in this last, we write
\b
\rho'=-\gamma\rho^2+\sigma\rho.
\label{eqr}
\e
Substituting this into the transformed version of equation (\ref{p3y}) and
looking
once again at dominant terms, we get
\b
\sigma(x)=-2\gamma Q(x)-\frac{\alpha+\gamma}{\gamma x}+\frac{\tau}{\rho}.
\e
Using this in the transformed equation yields a linear equation for $\rho$,
which has to be compatible with the Riccati equation
\b
\rho'=-\gamma\rho^2-\left(2\gamma Q(x)+\frac{\alpha+\gamma}{\gamma
x}\right)\rho
+\tau.
\label{p3ric}
\e
The analysis of the resulting compatibility condition depends on whether or not
$\tau$ is assumed to depend only on $x$ or if it is allowed to depend also on
$Q(x)$.
(We do not consider here the possible dependence of $\tau$ on $Q'(x)$.)
Assuming
$\tau=\tau(x,Q(x))$ and using the fact that $Q(x)$ satisfies
$P_{III}$, we obtain a polynomial in $Q'(x)$. The highest order coefficient
gives
\b
Q^2\tau_{QQ}-Q\tau_Q+\tau=0,
\e
whose general solution is given by
\b
\tau=\left[f_1(x)+f_2(x)\log Q\right]\,Q.
\e
If we now insert this form for $\tau$ in the compatibility condition and set to
zero
coefficients of the resulting polynomial in $Q$, $Q'$ and $\log Q$, we first
obtain
\ba
f_1(x) & = & \frac{a-\alpha}{\gamma x},\\
f_2(x) & = & 0,\\
c^2 & = & \gamma^2,\\
d^2 & = & \delta^2,
\ea
which means that the actual form of $\tau$ is
\b
\tau(x,Q)=\frac{a-\alpha}{\gamma x} Q.
\label{p3tau}
\e
Note that if we had taken $\tau$ to be a function of $x$ only, our
compatibility
condition would have led to $\tau=0$ and thus to $a=\alpha$, and we would only
have obtained restricted results. The main difference between the application
of
our method to $P_{III}$, and its previous application to $P_{II}$ and $P_{IV}$
\cite{GJP99}, is that for $P_{III}$, allowing $\tau$ to depend on $Q$ leads to
more general results, whereas for $P_{II}$ and $P_{IV}$ it does not.

Inserting the above results into the compatibility condition, we obtain
 from the next higher order term
the following shift between the parameters,
\b
2\gamma(b-\beta)+ba-\beta\alpha  =  0.
\label{beq}
\e
We now consider separately the remainder of our compatibility condition for the
two cases $b=0$ and $b\neq0$. We take first the case $b\neq0$. Solving
equation (\ref{beq}) for $a$ and substituting back into the compatibility
condition gives
the following additional constraints between the parameters,
\ba
(b-\beta)(b+\beta)(\gamma b-\delta\alpha-2\gamma \delta)
(\gamma b+\delta\alpha+2\gamma \delta) & = & 0, \label{4f} \\
(b-\beta)(b+\beta)(\gamma b-\delta\alpha-2\gamma \delta)
(\gamma b+\delta\alpha+2\gamma \delta)(2\gamma\beta+\beta\alpha-\gamma b) & = &
0.
\ea
Taking $b=\beta$ in  (\ref{4f}) just leads to the identity $y(x)=Q(x)$. However
the
remaining factors in  (\ref{4f}) lead to nontrivial B\"acklund transformations.
Taking $b=-\beta$
leads to the B\"acklund transformation
\ba
\rho& = & -2Q^2 \frac{A_1(x,Q,Q')}{A_2(x,Q,Q')},\label{bt1a}\\
\nonumber
A_1 & = & (2\gamma^2 x+\alpha\gamma x)Q'
+(\alpha\gamma^2 x+2\gamma^3 x)Q^2-(6\gamma^2+5\alpha\gamma+\alpha^2)Q\\
& & -\beta\gamma^2 x,\\
\nonumber
A_2 & = & \gamma^2 x^2Q'^2+2\gamma^3 x^2Q^2Q'-2\gamma^2 xQQ'+\gamma^4 x^2Q^4-
2\gamma^3 xQ^3\\
& & -(3\gamma^2+4\alpha\gamma+\alpha^2)Q^2-2\beta\gamma^2 x Q-\delta^2\gamma^2
x^2,\\
a & = & -\alpha-4\gamma,\\
b & = & -\beta,\\
c^2 & = & \gamma^2,\\
d^2 & = & \delta^2,\label{bt1b}
\ea
whereas taking $\gamma b \pm \delta (\alpha + 2\gamma)=0$ (which requires
$\delta\neq
0$) leads to
\ba
\rho & = & \mp\frac{(\gamma\beta\pm\delta\alpha\pm 2\gamma\delta)Q^2}
{\delta\left[\gamma xQ'+\gamma^2 xQ^2+(\alpha+\gamma)Q\pm\delta\gamma
x\right]},
\label{bt2a}\\
a & = & -\gamma\left(2\pm\frac{\beta}{\delta}\right),\\
b & = & \mp \left(2\delta+\frac{\alpha\delta}{\gamma}\right),\\
c^2 & = & \gamma^2,\\
d^2 & = & \delta^2.\label{bt2b}
\ea
Here the choice of sign of $\delta$ arises because of the way we have written
this
parameter in $P_{III}$.

If we now consider the case with $b=0$ and we use this in the remainder of our
compatibility condition, we recover the two  B\"acklund transformations above
(with $b=0$), and in addition
\ba
\rho & = & \frac{(a-\alpha)Q^2}{\gamma xQ'+\gamma^2 xQ^2+(\gamma+\alpha)Q},\\
b & = & \beta=0,\\
c^2 & = & \gamma^2,\\
d & = & \delta=0.
\ea
We note however that in the case $\beta=\delta=0$, $P_{III}$ is explicitly
solvable
\cite{FA,MCB,Gro}.

We now consider taking the opposite sign in front of the term in $\rho^2$ in
(\ref{eqr}), i.e.
\b
\rho'=\gamma\rho^2+\sigma\rho.
\e
However the results thus obtained can be written down simply by changing the
sign of
$\gamma$ in the results obtained above.

Thus far we have assumed $\gamma\neq 0$. However, since the change of variables
\b
y(x)=\frac{1}{m(x)}\label{rt}
\e
transforms $P_{III}$ in the variable $y(x)$ into $P_{III}$ in the variable
$m(x)$
but with parameters $\alpha'=-\beta$, $\beta'=-\alpha$, $\gamma'^2=\delta^2$
and
$\delta'^2=\gamma^2$, we can treat the case $\gamma=0$ as the case $\delta'=0$
after such a change of variables. This then requires that $\gamma'\neq0$, i.e.\
$\delta\neq0$. The remaining case $\gamma=\delta=0$ can be dealt with by
another
change of variables which maps $P_{III}$ with parameters $\alpha$, $\beta$,
$\gamma=0$
and $\delta=0$ onto $P_{III}$ with parameters $\alpha''=0$, $\beta''=0$,
$\gamma''=2\alpha$ and $\delta''=2\beta$ \cite{Gro} (and noting that if in
addition
$\alpha=0$ or $\beta=0$ the original copy of $P_{III}$ is in any case
explicitly
solvable \cite{FA,MCB,Gro}).

The four B\"acklund transformations obtained by taking
(\ref{bt2a})---(\ref{bt2b})
together with the possible change of sign $\gamma\rightarrow-\gamma$ correspond
to
the four fundamental B\"acklund transformations for $P_{III}$ in the case
$\gamma\delta\neq0$ (denoted $T_i$, $i=1,2,3,4$, in \cite{MCB}). All other
known
B\"acklund transformations for $P_{III}$ in this case $\gamma\delta\neq0$ can
be
expressed in terms of these $T_i$ together with (\ref{rt}) and simple
rescalings
\cite{MCB}.

The two B\"acklund transformations obtained by taking
(\ref{bt1a})---(\ref{bt1b})
together with the possible change of sign $\gamma\rightarrow-\gamma$ correspond
in
the case $\delta=0$ (and $\beta\gamma\neq0$) to a second iteration of a
B\"acklund
transformation given in \cite{Gro} (see also Theorem 4.1 in \cite{FA} for
$\gamma=0$,
or Transformation V in \cite{MCB}, for $\alpha\delta\neq0$) combined with
(\ref{rt})
and suitable rescalings. In the case
$\delta\neq0$ they correspond to the second iteration of appropriate
combinations
of (\ref{bt2a})---(\ref{bt2b}) (i.e. of the transformations $T_i$ in
\cite{MCB}).
However the general formulation of these second iterations presented here, of
B\"acklund transformations which are usually treated separately, appears to be
new.

\section{B\"acklund transformations for $P_{V}$}

We take $P_{V}$, with a similar renaming of paramaters as used for $P_{III}$,
in the form
\b
y''=\left(\frac{1}{2y}+\frac{1}{y-1}\right)y'^2
-\frac{y'}{x}+\frac{(y-1)^2}{2x^2}
\left(\alpha^2 y-\frac{\beta^2}{y}\right)+\frac{\gamma
y}{x}-\frac{\delta^2y(y+1)}{2(y-1)}.
\e
When applying the approach outlined above to $P_{V}$ in this form, the presence
of a
resonance at first order after the leading term
means that $\sigma$ [in an equation corresponding to (\ref{eqr})] is not
determined by a
linear algebraic equation. Rather
then continuing with the resulting set of equations thus obtained, we make
instead
the change of variables
\b
y(x)=\frac{m(x)}{m(x)-1}
\e
which gives
\ba
\nonumber
m(m-1)x^2m'' & = & \left(m-\frac{1}{2}\right)x^2m'^2+m(1-m)xm'+\delta^2x^2m^5\\
\nonumber
& & -\left(\gamma+\frac{5}{2}x\delta^2\right)xm^4+
 2(\gamma+\delta^2 x)xm^3 \\
& & +\frac{1}{2}\left(\beta^2-\alpha^2-\delta^2 x^2-2\gamma x\right)m^2
-\beta^2 m+
\frac{1}{2}\beta^2,
\label{eqm}
\ea
in which form no resonance at first order after the leading term
interferes with the determination of the corresponding $\sigma$.
A generic solution $m(x)$ of this last equation is transformed into another
solution $Q(x)$ of
the same equation (\ref{eqm}) but with
possibly different values of the parameters that we denote by $a, b, c$ and
$d$, as
\b
m(x)=\rho(x)+Q(x).
\label{r+q2}
\e
Substitution of (\ref{r+q2}) into equation (\ref{eqm}) gives an equation whose
dominant
terms near a pole of $\rho$, assuming $\delta\neq0$, are
\b
\delta^2 \rho^5+\rho\rho'^2\approx \rho^2\rho'',
\e
which can be integrated to give
\b
\rho'\approx \pm \delta \rho^2.
\e
We consider first the case with the plus sign in this last, and write
\b
\rho'=\delta \rho^2+\sigma \rho.
\label{eqr1}
\e
Substituting this into the transformed version of equation (\ref{eqm}) and
looking once again
at dominant terms, we get
\b
\sigma(x)=2\delta Q(x)-\frac{\delta+\gamma+\delta^2x}{\delta
x}+\frac{\tau}{\rho}.
\e
Using this in the transformed equation we obtain, as we did for the fourth
Painlev\'e
equation \cite{GJP99}, a quadratic in $\rho$ which has to be compatible with
the Riccati equation
\b
\rho'=\delta \rho^2 +\left(2\delta Q(x)-\frac{\delta+\gamma+\delta^2x}{\delta
x}\right)\rho+\tau.
\e
Again the analysis of the resulting compatibility condition depends on whether
or not $\tau$ is
assumed to depend only on $x$, or on both $x$ and $Q(x)$. [Again we do not
consider here the case
where $\tau$ is allowed to depend also on $Q'(x)$.] Assuming
$\tau=\tau(x,Q(x))$, and using the
fact that $Q$ satisfies $P_V$, we obtain a polynomial in $Q'$ whose higher
order coefficient
gives the following differential equation for $\tau$,
\b
2Q(2Q^3-4Q^2+3Q-1)\tau_{QQ}+(-4Q^3+6Q^2-1)\tau_Q+
2(2Q^2-2Q-1)\tau=0.
\e
This has general solution
\b
\tau=(2Q-1)\left[f_1(x)+f_2(x)\log\left(Q-\frac{1}{2}
+\sqrt{Q(Q-1)}\right)\right]+
2f_2(x)\sqrt{Q(Q-1)}.
\e
Inserting this form for $\tau$ into the compatibility condition we next obtain
\ba
f_1(x) & = & \frac{1}{2\delta x}(c-\gamma),\\
f_2(x) & = & 0, \\
d^2 & = & \delta^2,
\ea
which means that $\tau$ in fact takes the form
\b
\tau=\frac{1}{2\delta x}(c-\gamma)(2Q-1).
\e
 From this last we see that if we had assumed $\tau$ to be a function of $x$
only, then
as for $P_{III}$ we would only have obtained restricted results. Thus once
again we see
that allowing $\tau$ to depend on $Q$ leads to more general results.

Using the above results in our compatibility condition, we obtain from the next
higher
order term the following shift between
the parameters,
\b
a^2 =
\frac{\gamma^2-c^2+2\delta^2(\alpha^2+\beta^2-b^2)
+2\delta(\gamma-c)}{2\delta^2}.
\e
Using this in our compatibility condition then leads to
\ba\nonumber
b^2 & = &
\frac{4\beta^2\delta^3+2\left[(c+\gamma)\beta^2
+(c-\gamma)(\alpha^2-1)\right]\delta^2
+(\gamma^2-3c^2+2c\gamma)\delta-c^3+c\gamma^2}
{4\delta^2(c+\delta)}\\
& &
\ea
where we have assumed that $c+\delta\neq 0$ (the case $c+\delta= 0$ has to be
considered separately).
Substituting back once again then provides the additional condition
\b
(c-\gamma)(c+\gamma+2\delta)(c+\delta-\delta\alpha-\delta\beta)
(c+\delta+\delta\alpha-\delta\beta)(c+\delta-\delta\alpha+\delta\beta)
(c+\delta+\delta\alpha+\delta\beta)=0.
\e
This provides essentially three different cases to consider, since the last
three factors
are related to the third under changes of sign of $\alpha$ and $\beta$.
These three cases are:
\ba
c& = & \gamma,\\
c & = & -\gamma-2\delta,\\
c & = & \delta(\beta+\alpha-1).
\ea
The case $c=\gamma$ corresponds to the identity transformation (this is easily
seen
by substituting this condition into the above expressions for $a^2$ and $b^2$.)
The other
two cases lead to two nontrivial B\"acklund transformations, i.e.
\ba
\rho& = & Q(Q-1) \frac{A_1(x,Q,Q')}{A_2(x,Q,Q')},\label{bt3a}\\
\nonumber
A_1 & = & -2\delta x(\delta+\gamma)Q'+2\delta^2 x(\delta+\gamma)Q^2+
2(\delta^2-\delta^3x+2\gamma\delta-\delta^2\gamma x+\gamma^2)Q\\
& & +\delta^2(\beta^2-\alpha^2-1)-2\gamma\delta-\gamma^2,\\
\nonumber
A_2 & = &
\delta^2x^2Q'^2+2\delta^3x^2(Q-Q^2)Q'+\delta^4x^2Q^4
-2\delta^4x^2Q^3-\alpha^2\delta^2
\\
& &
+(\delta^4x^2-\delta^2-2\gamma\delta-\gamma^2)Q^2
+(\delta^2+\alpha^2\delta^2-
\beta^2\delta^2+2\gamma\delta+\gamma^2)Q,\\
a^2 & = & \beta^2,\\
b^2 & = & \alpha^2,\\
c & = &  -\gamma-2\delta,\\
d^2 & = & \delta^2,\label{bt3b}
\ea
and
\ba
\rho& = & \frac{2Q(Q-1)(\alpha\delta+\beta\delta-\gamma-\delta)}
{2\delta
xQ'-2\delta^2xQ^2+2(\gamma+\delta+\delta^2x)Q
+\delta(\alpha-\beta-1)-\gamma},\label{bt4a}\\
a^2 & = & \left[\frac{\gamma+\delta(\alpha-\beta+1)}{2\delta}\right]^2,\\
b^2 & = & \left[\frac{-\gamma+\delta(\alpha-\beta-1)}{2\delta}\right]^2,\\
c & = &  \delta(\beta+\alpha-1),\\
d^2 & = & \delta^2.\label{bt4b}
\ea
We also have of course B\"acklund transformations obtained from the above under
combinations of $\alpha\rightarrow-\alpha$, $\beta\rightarrow-\beta$.
As mentioned earlier, we also have our compatibility condition for the case
$c+\delta=0$,
which needs to be considered separately. However this leads only to restricted
cases of the above B\"acklund transformations.

We now consider taking the opposite sign in front of the term in $\rho^2$ in
(\ref{eqr1}), i.e.
\b
\rho'=-\delta \rho^2+\sigma \rho.
\e
The results thus obtained are as above with $\delta\rightarrow-\delta$.

In the above we have assumed $\delta\neq0$. We note that $P_V$ in the case
$\delta=0$
can be reduced to $P_{III}$ with $\gamma\delta\neq0$ \cite{Gro}, which we
considered in
the previous section.

The B\"acklund transformations represented by (\ref{bt4a})---(\ref{bt4b}) do
not seem to have
been given before. They correspond to a composition of what appear to be the
only previously
known (see \cite{FA,Gro}) auto-B\"acklund transformations for $P_V$ in the case
$\delta\neq0$.
These known B\"acklund transformations, as well as this composistion, are
derived in our approach
by consideration of the ODE satisfied by $\rho(x)$ (we leave details of this
derivation to
\cite{GJP00}; the resulting
discussion is analogous
to our discussion of $P_{IV}$ in \cite{GJP99}). The B\"acklund transformations
represented by
(\ref{bt3a})---(\ref{bt3b}) correspond to second iterations, with appropriate
choices of signs, of (\ref{bt4a})---(\ref{bt4b}).

\section{Conclusions}

We have applied our approach to obtaining B\"acklund transformations, which is
based
on mappings preserving natural subsets of movable poles, to $P_{III}$ and
$P_V$. For
$P_{III}$ we have recovered the four fundamental B\"acklund transformations for
the
case $\gamma\delta\neq0$, as well as what appears to be a new general
formulation of
the second iteration of these or of a further B\"acklund transformation which
exists
in the case $\gamma\delta=0$. For $P_V$ our approach allows us to recover what
appear to be all known
B\"acklund transformations (details in \cite{GJP00}), and in addition a
composition of these
which seems not to have been given previously. A crucial point in obtaining
such general results for
$P_{III}$ and $P_V$ is our allowing $\tau$ to depend not only on $x$ but also
on $Q(x)$.
Further aspects of our approach for $P_{III}$ and $P_V$ are considered in
\cite{GJP00}.

\section*{Acknowlegements}

The research of PRG was supported in part by the DGICYT under contract
PB95-0947. NJ
acknowledges support from the Australian Research Council. AP thanks the
Ministry
of Education and Culture of Spain for a post-doctoral fellowship.

\section*{Appendix}

In this appendix we briefly compare the approach outlined here with the
standard
Painlev\'e truncation. We take as an example the third Painlev\'e equation
$P_{III}$.

Seeking a solution of (\ref{p3y}) in the form of a so-called truncated
Painlev\'e
expansion, we obtain
\b
y=\frac{\varphi'}{\gamma\varphi}
-\frac{1}{2\gamma}\left(\frac{\varphi''}{\varphi'}+\frac{\alpha+\gamma}{\gamma
x}\right),
\label{tr1}
\e
together with an equation of the form $A\varphi^{-1}+B=0$, for some $A$, $B$.

The truncated expansion (\ref{tr1}) can be compared with the results obtained
in Section 2 by linearising the Riccati equation (\ref{p3ric}), (\ref{p3tau})
by setting
$\rho=\varphi'/(\gamma\varphi)$, which then leads to
\b
Q=-\frac{1}{2\gamma-\frac{(a-\alpha)\varphi}{x\varphi'}}
\left(\frac{\varphi''}{\varphi'}
+\frac{\alpha+\gamma}{\gamma x}\right),
\e
and thus (\ref{r+q}) becomes
\b
y=\frac{\varphi'}{\gamma\varphi}
-\frac{1}{2\gamma-\frac{(a-\alpha)\varphi}{x\varphi'}}
\left(\frac{\varphi''}{\varphi'}
+\frac{\alpha+\gamma}{\gamma x}\right).
\e
This is the same as the truncated Painlev\'e expansion (\ref{tr1}) only in the
case
$a=\alpha$, and thus we see that the only results obtainable using (\ref{tr1})
will
be those for this restricted case. That is, we can only obtain the identity or
restricted cases of the B\"acklund transformations presented in Section 3.
Deriving
these results requires applying the method presented in \cite{jp1,jp2} rather
than
a standard truncation approach.

Similar remarks hold for $P_{IV}$ and $P_{V}$, or in general when the function
$\tau\neq0$.

\end{document}